# TOWARDS RESPONSIBLE QUANTUM TECHNOLOGY

SAFEGUARDING, ENGAGING AND ADVANCING QUANTUM R&D

Mauritz Kop, Mateo Aboy, Eline De Jong, Urs Gasser, Timo Minssen, I. Glenn Cohen, Mark Brongersma, Teresa Quintel, Luciano Floridi, Ray Laflamme*

## ABSTRACT

The expected societal impact of quantum technologies (QT) urges us to proceed and innovate responsibly. This article proposes a conceptual framework for Responsible QT that seeks to integrate considerations about ethical, legal, social, and policy implications (ELSPI) into quantum R&D, while responding to the *Responsible Research and Innovation* dimensions of anticipation, inclusion, reflection and responsiveness. After examining what makes QT unique, we argue that quantum innovation should be guided by a methodological framework for Responsible QT, aimed at jointly *safeguarding* against risks by proactively addressing them, *engaging* stakeholders in the innovation process, and continue *advancing* QT ('SEA'). We further suggest operationalizing the SEA-framework by establishing quantum-specific guiding principles. The impact of quantum computing on information security is used as a case study to illustrate (1) the need for a framework that guides Responsible QT, and (2) the usefulness of the SEA-framework for QT generally. Additionally, we examine how our proposed SEA-framework for responsible innovation can inform the emergent regulatory landscape affecting QT, and provide an outlook of how regulatory interventions for QT as base-layer technology could be designed, contextualized, and tailored to their exceptional nature in order to reduce the risk of unintended counterproductive effects of policy interventions.

Laying the groundwork for a responsible quantum ecosystem, the research community and other stakeholders are called upon to further develop the recommended guiding principles, and discuss their operationalization into best practices and real-world applications. Our proposed framework should be considered a starting point for these much needed, highly interdisciplinary efforts.

## TABLE OF CONTENTS

# I. INTRODUCTION

More than a hundred years ago scientists discovered that the world at a very small scale behaves very differently from what we are used to in our daily lives. In its inaugural century, quantum science primarily concentrated on understanding the rules and principles that govern physical reality at the scale of atoms. During the first quantum revolution, the theory of quantum mechanics was developed and experimentally validated. The resulting quantum mechanical principles were then used to create first-generation (1G) quantum technologies (QT) such as transistors, lasers, and MRI. More recently, the rapid advances in nanotechnology, optics, high performance computer engineering, and communications have unfolded a myriad of new ways to measure, control, and utilize the quantum properties of light and matter.

We are currently witnessing a second quantum revolution, where quantum mechanical principles and 1G QT are employed to realize a second generation (2G) of quantum technologies. This generation of technologies *directly harness* quantum mechanical phenomena such as superposition, entanglement, and tunneling (Box 1).[1] The resulting 2G QT highlight the potential for quantum information science to develop into quantum technologies across several domains.[2]

Applications of 2G QT include 1) simulating quantum systems to enhance our fundamental understanding of nature and its applications,[3] such as modelling chemical processes in drug development, 2) achieving unprecedented precision in measurement through quantum sensing and metrology,[4] such as Rydberg atom sensors and atomic clocks, 3) solving mathematical and computational problems beyond the reach of classical computing by using quantum computers to formulate and deploy quantum algorithms that leverage quantum superposition and entanglement,[5] and 4) constructing a new generation of secure communication systems.[6]

---

\* Mauritz Kop is TTLF Fellow and Visiting "Quantum & Law" Scholar at Stanford Law School, Stanford University, Stanford, CA, USA, and Director at AIRecht.nl, The Netherlands, mkop@stanford.edu, Mateo Aboy is Principal Research Scholar in Innovation, AI & Law at the LML, University of Cambridge, UK, and Affiliated Professor and Fellow at the Centre for Advanced Studies in Biomedical Innovation Law (CeBIL), University of Copenhagen, Denmark, ma608@cam.ac.uk, Eline De Jong is PhD candidate Philosophy and Ethics of Quantum-Safe Cryptography at the Institute for Logic, Language and Computation and the Institute of Physics at the University of Amsterdam, The Netherlands, e.l.dejong@uva.nl, Urs Gasser is Dean of the TUM School of Social Sciences and Technology at the Technical University of Munich, Germany, urs.gasser@tum.de, Timo Minssen is Professor of Law and the Founding Director of the Center for Advanced Studies in Biomedical Innovation Law (CeBIL), University of Copenhagen, Denmark. He is also a LML Research Affiliate at the University of Cambridge, UK, timo.minssen@jur.ku.dk, I. Glenn Cohen is Deputy Dean and James A. Attwood and Leslie Williams Professor of Law, Harvard Law School, and Faculty Director, Petrie-Flom Center for Health Law Policy, Biotechnology & Bioethics, MAS, USA, igcohen@law.harvard.edu, Mark Brongersma is the Stephen Harris Professor of Engineering at Stanford University. He is also a Fellow of the OPTICA, the SPIE, and the American Physical Society, Stanford, USA, brongersma@stanford.edu, Teresa Quintel is Assistant Professor at the European Centre on Privacy and Cybersecurity (ECPC) at Maastricht University, The Netherlands, t.quintel@maastrichtuniversity.nl, Luciano Floridi is Professor of Philosophy and Ethics of Information at the University of Oxford and Professor of Sociology of Culture and Communication at the Alma Mater Studiorum University of Bologna, where he directs the Centre for Digital Ethics, Oxford, UK, luciano.floridi@oii.ox.ac.uk, Ray Laflamme is the Mike and Ophelia Lazaridis "John von Neumann" Chair in Quantum Information at the Institute for Quantum Computing at the University of Waterloo and Associate Faculty at the Perimeter Institute for Theoretical Physics in Waterloo, Canada, laflamme@uwaterloo.ca.

[1] *See e.g.,* John Preskill, *Quantum computing 40 years later*, arXiv preprint arXiv:2106.10522, 2021.
[2] *See* J. P. Dowling, G. J. Milburn, *Quantum Technology: The Second Quantum Revolution*, PHILOSOPHICAL TRANSACTIONS OF THE ROYAL SOCIETY OF LONDON. SERIES A: MATHEMATICAL, PHYSICAL AND ENGINEERING SCIENCES VOLUME 361, ISSUE 1809 (2003). https://doi.org/10.1098/rsta.2003.1227
[3] *See e.g.,* Pouse, W., Peeters, L., Hsueh, C.L. *et al. Quantum simulation of an exotic quantum critical point in a two-site charge Kondo circuit,* NAT. PHYS. (2023), https://doi.org/10.1038/s41567-022-01905-4
[4] *See e.g.,* C. L. Degen, F. Reinhard, P. Cappellaro, *Quantum sensing*, REV. MOD. PHYS. 89, 035002 (2017), https://doi.org/10.1103/RevModPhys.89.035002

During the early pioneering years of scientific discovery, there was no imminent need for researchers to engage directly with the ethical, legal, social, and policy implications of QT (Quantum-ELSPI). But as we see 2G QT move from pure science to application in real-world we must broaden our lens to considerer the development and use of QT in human and societal contexts. What will it mean for the law and other societal institutions? How should QT be developed and regulated?[7] The introduction of 2G QT in society raises important legal and regulatory questions pertaining to national and economic security, dual use, privacy, product safety and liability, intellectual property, fair competition, and equality. For example, quantum algorithms have the potential to break current cryptography protocols, threatening the information security and data privacy of its users, thereby destabilizing society and undermining trust in its institutions. In this paper we use information security as the illustrative study of the challenges and the application of our proposed framework for Responsible QT.

The current breadth, speed of maturation, and potential impact of 2G QT in human and societal contexts make it an urgent priority to engage with the emerging interdisciplinary research field of Quantum-ELSPI.[8] This approach can help to guide R&D and application of QT into a modus that is ethical and socio-economically sustainable, while also promoting responsible technological advancement and innovation. Steering towards beneficial societal outcomes, we propose a conceptual framework for Responsible QT that is informed by Quantum-ELSPI considerations and responds to key Responsible Research and Innovation (RRI) dimensions.

The article is structured as follows. Part II discusses what makes QT unique and defines quantum mechanical effects such as superposition, entanglement, tunneling, and quantization. Part III conceptualizes the Responsible QT paradigm and provides arguments as to why we need to proactively fill the current responsibility gap. After explaining RRI in terms of Quantum-ELSPI, Part IV then argues that quantum innovation should be guided by a framework for Responsible QT, aimed at jointly *safeguarding* against risks by proactively addressing them, *engaging* stakeholders in the innovation process, and continue *advancing* QT ('SEA'). We further suggest operationalizing the SEA-framework by establishing ten quantum-specific guiding principles. Part V illustrates the importance of such practices by examining the example of information security in the post-quantum era, recommending that research on and development of QT should be accompanied by risk-based quantum impact assessments focused on information security risks and implementing quantum-safe information security controls to mitigate such risks. Part VI analyses how our proposed SEA-framework for Responsible QT can inform the emergent regulatory landscape affecting QT, taking as examples the two recent Executive Orders signed May 4 2022 by President Biden, and the Quantum Computing Cybersecurity Preparedness Act that became public law on December 21 2022. Laying the groundwork for a responsible, values-based quantum ecosystem, the conclusion calls upon the collaboration of multidisciplinary teams of diverse quantum stakeholders to discuss and orchestrate normative dimensions of QT futures, and pathways to build towards them.

---

[5] *See e.g.,* Bohr, N. *The Quantum Postulate and the Recent Development of Atomic Theory*, NATURE 121, 580–590 (1928). https://doi.org/10.1038/121580a0. *See also* Caltech, What Is Superposition and Why Is It Important? https://scienceexchange.caltech.edu/topics/quantum-science-explained/quantum-superposition, and L. Billings, *Explorers of Quantum Entanglement Win 2022 Nobel Prize in Physics*, SCIENTIFIC AMERICAN 2022, https://www.scientificamerican.com/article/explorers-of-quantum-entanglement-win-2022-nobel-prize-in-physics1/

[6] *See e.g.,* M. Aboy, T. Minssen, M. Kop, *Mapping the Patent Landscape of QT: Patenting Trends, Innovation and Policy Implications*, INTERNATIONAL REVIEW OF INTELLECTUAL PROPERTY AND COMPETITION LAW (IIC), VOLUME 53, PP. 853-882, SPRINGER NATURE, (2022). https://link.springer.com/article/10.1007/s40319-022-01209-3.

[7] A pioneering take-up of this challenge is Project Q, an initiative at the University of Sydney, Australia, aimed at investigating the geopolitical and societal implications of quantum innovation in computing, communications and artificial intelligence, *see* https://projectqsydney.com/ *See also* C. J. Hoofnagle, S. Garfinkel, *Law and Policy for the Quantum Age* (BERKELEY, 2021), pp 303-456.

[8] For an explanation of key Quantum-ELSPI elements including a selection of relevant Quantum-ELSPI questions, *see* M. Kop, *Quantum ELSPI: Ethical, Legal, Social and Policy Implications of Quantum Technology*, DIGITAL SOCIETY (SPRINGER NATURE), July 28, 2021, https://law.stanford.edu/publications/quantum-elspi-ethical-legal-social-and-policy-implications-of-quantum-technology/.

## II. WHAT MAKES QUANTUM TECHNOLOGIES UNIQUE?

For the last 70 years, the transistor has been the fundamental building block in our electronic devices to enable complex computations and information manipulations. It has had a profound impact on the way we work, interact, and think. The rules of how information can be manipulated with classical devices are ingrained in classical physics, which limits the precision of sensors and the problems we can solve efficiently with classical computers. This latter part is encoded in the Strong Church-Turing thesis at the basis of today's classical computer science.[9] This thesis stipulates that all computers are born (roughly) equal because any real-world computation can be translated into an equivalent computation involving a Turing machine, i.e. if a problem is a hard one and thus requiring an exponential amount of computational resources, it will be hard for all computers.[10] For more than 80 years this fundamental tenet was accepted as true, and it is this tenet that QTs challenge.[11] The technology community was surprised when it was demonstrated that quantum mechanics for computing could solve problems for which we do not have efficient classical algorithms such as finding the prime factors of large composite integers.[12] These computationally hard problems for classical computers are the basis of our public-key cryptography infrastructure that secures internet communications, but quantum algorithms have been discovered that could solve these classically intractable problems (e.g., Shor's algorithm for prime factorization).[13]

In general, second generation (2G) QT *directly harness* quantum mechanical phenomena such as quantum superposition, entanglement, and tunneling (discussed in Box 1) to achieve *quantum advantage* over state-of-the-art technologies in both qualitative and quantitative ways.[14] This includes deploying quantum algorithms for classically intractable problems, simulation of quantum systems beyond the capabilities for our most powerful classical supercomputers, quantum sensing that can sense weak forces due to higher sensitivity and spatial resolution unachievable by classical sensors, and secured quantum communications. In other words, QT take a fundamentally novel approach to computation, sensing and communication, coming with potentially disruptive effects. Imagine the consequences of a first mover being able to crack our ubiquitous RSA encryption and similar public-key cryptosystems in a few seconds by implementing Shor's algorithm in a sufficiently fault-tolerant quantum computer. QT are categorically different from the technological improvements that we have seen over the last 70 years because of the nature of the advantages that *quantum dominance* could entail.

---

[9] For an accessible explanation of the Strong Church-Turing thesis for a broader public, *see* Copeland, B. Jack, "The Church-Turing Thesis", *The Stanford Encyclopedia of Philosophy* (Summer 2020 Edition), Edward N. Zalta (ed.), https://plato.stanford.edu/archives/sum2020/entries/church-turing/

[10] For a selection of the original Church-Turing papers, *see* Alan Turing, *On computable numbers, with an application to the Entscheidungsproblem*, PROCEEDINGS OF THE LONDON MATHEMATICAL SOCIETY, VOLUME 42, ISSUE 2, PP 230-265, 2 Nov 1936, http://140.177.205.52/prizes/tm23/images/Turing.pdf and Church, A. *Abstract No. 204*. BULL. AMER. MATH. SOC. 41, 332-333, 1935.

[11] For a technical overview of the history of the Strong Church-Turing thesis, *see* Robert I. Soare, *Turing oracle machines, online computing, and three displacements in computability theory*, ANNALS OF PURE AND APPLIED LOGIC, VOLUME 160, ISSUE 3, 2009, pp 368-399, ISSN 0168-0072, https://doi.org/10.1016/j.apal.2009.01.008

[12] For further advanced reading on quantum mechanics for computing, *see* David Deutsch, *Quantum theory, the Church-Turing principle and the universal quantum computer*, in PROCEEDINGS OF THE ROYAL SOCIETY OF LONDON A 400, pp. 97-117 (1985), (*Communicated by R. Penrose, F.R.S. — Received 13 July 1984*), https://www.daviddeutsch.org.uk/wp-content/deutsch85.pdf

[13] *See* P. Shor, *Polynomial-Time Algorithms for Prime Factorization and Discrete Logarithms on a Quantum Computer* SIAM J.SCI.STATIST.COMPUT. 26 1484 (1997).

[14] For further reading -for a general audience- on the background of these quantum mechanical phenomena, see Chris Bernhardt, *Quantum Computing for Everyone*, Sept. 8, 2020, THE MIT PRESS, https://mitpress.mit.edu/9780262539531/quantum-computing-for-everyone/; Feynman, R. P., Leighton, R.B., Sands, M.

---

**Box 1: Superposition, Entanglement, Tunneling, and Quantization in QT**

**Superposition**: a concept in quantum mechanics that states that a particle can exist in multiple states at the same time. Superposition is directly harnessed in quantum computing to achieve *quantum advantage* over classical computation. In a classical computer, data is represented as bits, which can have one of two states 0 or 1. In a quantum computer, quantum information is unfolded by quantum bits (qubits), which can be in a $|0>$ state or in a $|1>$ as well, but also in a superposition of both basis states as a linear combination $|\psi> = \alpha|0> + \beta|1>$ where $\alpha$ and $\beta$ are complex numbers corresponding to probability amplitudes. Thus, a quantum computer consisting of $n$ qubits can exist in a superposition of $2^n$ states enabling unprecedented parallel computation.

**Entanglement**: a phenomenon in which two or more particles interact in such a way as to become mutually dependent on one another, even when separated by great distances. When two systems are entangled there exists a special connection between them. If two qubits are entangled, it means that, on measurement the results are strongly correlated even if the qubits are physically separated across great distances. For instance, if the first qubit is measured to be in state $|0>$, the second entangled qubit will also be found to be in state $|0>$. Entanglement is directly harnessed in quantum computing along with superposition to achieve quantum parallelism resulting in quantum algorithms with exponential speed-up over classical computations. It is also harnessed in quantum communications by taking advantage of unique correlations exhibited by entangled qubits and quantum cryptography, particularly quantum key distribution (QKD).

**Tunneling**: the ability of quantum systems to go across an energy barrier. A phenomenon in which a particle can pass through a barrier that it does not have enough energy to surmount. Tunneling is directly harnessed in quantum annealing/adiabatic computation to solve quantum simulation and optimization problems. It is also used in applications such as scanning tunneling microscopes, tunneling diodes, and quantum sensors. As the transistors (MOSFETS) used in our classical computers get smaller, they also exhibit quantum mechanical tunnelling from source to gate oxide due to the thickness of the oxide layers and quantum mechanical tunnelling from source to drain when the channel lengths are less than 10 nm. While in classical technologies at the small-scale such as MOSFETS quantum effects are often sources of *imperfections to avoid,* in 2G QT these quantum mechanical effects are *directly harnessed* to achieve *quantum advantage.*

**Quantization**: a fundamental principle of quantum mechanics that is used to describe the behavior of physical systems at the atomic and subatomic level. The allowed energies of a tightly confined system of particles at quantum scales are restricted to a discrete set. It is an essential part of many technologies including lasers and the physical realization of qubits no matter which qubit approach one takes, such as trapped ions, atoms, photons, quasi particles, or superconducting oscillator circuits. In order to function as a qubit, the discrete energy levels of the superconducting circuit must be carefully controlled and manipulated.

---

(1965). *The Feynman Lectures on Physics*, VOLUME 3, ADDISON-WESLEY, READING, MA, https://www.feynmanlectures.caltech.edu/III_01.html#Ch1-S8; and Phillip Kaye, Raymond Laflamme & Michele Mosca*, An Introduction to Quantum Computing,* Jan 18, 2007, OXFORD UNIVERSITY PRESS.

## III. CONCEPTUALIZING RESPONSIBLE QUANTUM TECHNOLOGY

We posit that the exceptional nature of QT demands proactive integration of ELSPI considerations throughout the entire QT R&D lifecycle. The potential game changing character of QT, unlocking novel approaches in both research and innovation, comes with the expectation that it will affect our world in a myriad of ways.[15] Pertinent examples of these challenges concern national and economic security, dual use, privacy, product safety and liability, intellectual property, fair competition, and equality.

While we should build upon successes and failures from dealing with -and establishing responsible technology frameworks for- closely related fields such as AI, nanotechnology, biosciences, semiconductors, and nuclear,[16] the exceptionality of 2G QT and their expected societal impact including the nature of the advantages that *quantum dominance* could cause, demand a tailored approach.[17]

It is this expected impact, stemming from the fundamentally novel qualities of QT itself[18], that urges us to proceed *responsibly*.[19] A first key step thus is a reflection on that very concept: What does responsible research and innovation amount to in the context of QT? A conceptualization of Responsible QT serves as a touchstone for the nascent Quantum-ELSPI domain.

Founded on the concept of Responsible QT, we call for integrating ELSPI considerations within quantum innovation, taking the societal context into account early on. We propose an overarching framework for responsible quantum innovation (See Fig. 1) and provide suggestions for its operationalization by establishing quantum-specific guiding principles.[20] Our proposition is aimed at researchers, developers, innovators, and regulators, but it may also inspire other stakeholders. The proposed Responsible QT approach should be considered as a starting point for much needed highly interdisciplinary efforts (Box 2).

The concept of Responsible QT is aimed at ensuring that ethical, legal, socio-economic, societal, and philosophical dimensions are identified and discussed while QTs are still shapeable.[21] From a normative perspective, the objective is to capture and promote beneficial opportunities expected from quantum innovation while managing potential downside risk by putting into place technical, organizational, and policy measures appropriate to the risk. At present, many QT applications, such as large-scale fault-tolerant quantum computers[22] or the Quantum Internet are still in the basic research stage, and indeed many societal implications remain unknown.[23] Other applications, such as noisy intermediate-scale quantum (NISQ) computers, and *in silico* design of new catalysts, materials, and pharmaceuticals by using scalable quantum simulation of molecular energies, are at higher technology readiness levels (TRL).[24] Exactly because of the early stage of the technology and its far-reaching potential, we have a shared opportunity and responsibility to shape its development toward desirable societal outcomes.[25] Responsible QT urges that QT are designed, produced, marketed, and protected in an ELSPI-sensitive manner. The need for Responsible QT becomes strikingly evident when examining prospective trajectories where QT software and hardware structures are developed and commercialized without such considerations. This includes use cases pertaining to national defense and commercial security threats, unwanted disclosure of trade and state secrets, large scale privacy loss, and winner-takes-all dynamics. The next section explores how one could fill these identified responsibility gaps.

---

[15] *See e.g.,* Stephen Witt, The World-Changing Race to Develop the Quantum Computer, The New Yorker, Dec 12, 2022, https://www.newyorker.com/magazine/2022/12/19/the-world-changing-race-to-develop-the-quantum-computer

[16] *See e.g.,* International Atomic Energy Agency, *Establishing a Code of Ethics for Nuclear Operating Organizations*, IAEA NUCLEAR ENERGY SERIES NO. NG-T-1.2, IAEA, VIENNA (2007).

[17] This is the main reason why we cannot just apply methodological frameworks for AI, biosciences, nuclear fission or nanotechnology to quantum technology – these use cases are categorically different in many dimensions. We should however transplant the parts of those existing frameworks that are relevant to, or of special value for QT use cases, e.g. the parts that address dual use characteristics, or that apply to all general purpose technologies.

[18] As we explain in box 1 and section 2, QT radically differ from other technologies that rely on, and harness classical physics. The novel nature of QT calls for a fundamental reflection on the direct implications for how we approach ethics, law and policy – which is in its core a philosophical exercise. In this paper, however, we address the novel nature of QT indirectly, by focusing on the expected societal impact that flows from it.

[19] In parallel, the free world should prioritize environmental, social, and governance (ESG) investing in QT R&D, to avoid losing the competition for technological supremacy from countries with incompatible ideologies.

[20] *Compare to* P. Inglesant, M. Jirotka & M. Hartswood, *Responsible Innovation in QT applied to Defence and National Security*, NQIT, (2018).

[21] For further reading on the ethical, legal, socio-economic, societal, and philosophical dimensions of QT, *see* M. Kop, *Ethics in the Quantum Age*, PHYSICS WORLD, DEC., 31 (2021), https://physicsworld.com/a/why-we-need-to-consider-the-ethical-implications-of-quantum-technologies/.

## IV. A FRAMEWORK FOR RESPONSIBLE QUANTUM INNOVATION

Responsible Research and Innovation (RRI) is an approach that aims to ensure that scientific and technological developments are carried out in a way that is socially desirable, ethically acceptable, and sustainable.[26] RRI is a process that involves a continuous dialogue between researchers, citizens, industry, policy makers, and other stakeholders, in order to actively anticipate and assess the potential social and ethical implications of research, development and innovation.

The concept of RRI emerged as a response to the growing recognition that scientific and technological progress can have unintended negative consequences for society and should be guided by ethical and social considerations. RRI aims to integrate these considerations, norms, and values, into all stages of the R&D and innovation process, from the design of research projects to the dissemination of results.

In accordance with the European Commission, four important dimensions of RRI are anticipation, inclusion, reflection, and responsiveness.[27] Anticipation entails identifying and addressing potential social and ethical issues that may arise. Inclusion implies involving a wide range of stakeholders in the innovation process, including those who may be affected by the outcomes. Reflection concerns ongoing evaluation of the values and assumptions that underpin research and innovation, and considering how they may influence the outcomes. Responsiveness implicates reacting to the feedback and concerns of stakeholders and adapting accordingly. In sum, RRI is a way of ensuring that research, development, and innovation are aligned with the needs and values of society, and that they contribute to the common good.

---

[22] *See e.g.,* M. Brooks, *What's next for quantum computing*, MIT TECHNOLOGY REVIEW, (2023), https://www.technologyreview.com/2023/01/06/1066317/whats-next-for-quantum-computing/

[23] *See* on social implications of QT in connection to EDI-frameworks (Equity, Diversity, and Inclusion), G. Wolbring, *Auditing the 'Social' of Quantum Technologies: A Scoping Review,* SOCIETIES, 12(2), 41. (2022), https://doi.org/10.3390/soc12020041

[24] *See e.g.,* P. O'Malley *et al., Scalable Quantum Simulation of Molecular Energies*, PHYS. REV. X 6, 031007 (2016), https://journals.aps.org/prx/abstract/10.1103/PhysRevX.6.031007; and Simson L. Garfinkel and Chris J. Hoofnagle, *ACM TechBrief: Quantum Computing and Simulation*, ASSOCIATION FOR COMPUTING MACHINERY, NEW YORK, NY, USA (2022).

[25] E. De Jong, *Own the Unknown: An Anticipatory Approach to Prepare Society for the Quantum Age,* DIGITAL SOCIETY, QUANTUM-ELSPI TC, 1, SPRINGER NATURE, (2022), https://link.springer.com/article/10.1007/s44206-022-00020-4 Topical Collection: https://link.springer.com/collections/eiebhdhagd

[26] For framing and operationalizing RRI into stakeholder's everyday practice, *see* https://rri-tools.eu/research-community

[27] This has been referred to as the AIRR framework. *See* J. Stilgoe, R. Owen, and P. Macnaghten, *Developing a Framework for Responsible Innovation*, RESEARCH POLICY 42 (9) (2013): 1568–1580, https://doi.org/10.1016/j.respol.2013.05.008

In response to these key RRI dimensions, we argue that Responsible QT calls for quantum innovation that proactively addresses risks, takes emerging challenges heads-on, and seizes the opportunities that arise with QT.[28] This results in what we call the 'SEA-framework for Responsible Quantum Technology', aimed at *safeguarding* against risks, *engaging* stakeholders in the innovation process and *advancing* QT.[29] Safeguarding, engaging, and advancing QT in unison constitutes the triumvirate of Responsible QT. These objectives can be pursued at three, interrelating levels: (1) the technical level, (2) the ethical level, and (3) the societal, legal and policy level (Fig. 1).[30] The *technical* level involves setting standards, accountability, and governance mechanisms for QT. The *ethical* level provides the foundational principles and criteria that should guide both the technical controls as well as the QT *societal, legal and policy* level, which focuses on the societal impacts of legal frameworks and related policy decisions. Other technologies, including nuclear, biotech, and artificial intelligence (AI) have shown the complexity of developing principles, standards, and regulatory frameworks.[31] Despite its unique characteristics and potential impact, such effort has not been undertaken for QT yet. The methodological framework proposed here seeks to provide a starting point for this highly interdisciplinary endeavor.

Before elaborating on the SEA-framework, we should reflect on three issues in advance. *First*, as with any ethical framework, categories and principles can conflict. This is inherent to their value ladenness and there is no panacea to avoid tensions. For example, deontological (defining the moral course of action take in terms of right and duties), utilitarian (defining the right in terms of maximizing the good, and in the instance of classic utilitarianism the greatest happiness irrespective of its distribution and understood in terms of pleasure and pain) and consequentialist (justifying actions in the context of results) approaches in ethical decision-making may inform different outcomes.[32] Instead of considering this a weakness of the framework, such tensions reflect the complexity of the real world. Possible conflicts should thus lead to explicit discussions about how to balance different aims and stakes.

*Second*, the envisioned principles are not necessarily exclusive to QT. By 'Principles for Responsible Quantum Technology' we aim at principles that adequately respond to the SEA-objectives in the context of QT. The resulting set can thus consist of a mix of quantum specific principles and more general principles that are particularly relevant to quantum innovation.[33]

---

[28] For further reading on the 6 RRI themes (public engagement, open access, gender equality, ethics, science education, and governance) in the context of quantum technologies, *see* Ten Holter, C., Inglesant, P., & Jirotka, M. *Reading the road: challenges and opportunities on the path to responsible innovation in quantum computing*, TECHNOLOGY ANALYSIS AND STRATEGIC MANAGEMENT, (2021), https://doi.org/10.1080/09537325.2021.1988070; and Inglesant, P., Ten Holter, C., Jirotka, M., & Williams, R. *Asleep at the wheel? Responsible Innovation in quantum computing,* TECHNOLOGY ANALYSIS AND STRATEGIC MANAGEMENT, 0(0), 1–13. (2021), https://doi.org/10.1080/09537325.2021.1988557.

[29] A 'sea change' refers to a profound or notable transformation. *Compare to* the 'AREA Framework - Anticipate, Reflect, Engage, Act', which aims to embed RRI practices into emerging technologies, *see* R. Owen, J. Stilgoe, and P. Macnaghten, *A framework for responsible innovation,* RESPONSIBLE INNOVATION: MANAGING THE RESPONSIBLE EMERGENCE OF SCIENCE AND INNOVATION IN SOCIETY, 31 (2013): 27-50, https://onlinelibrary.wiley.com/doi/10.1002/9781118551424.ch2 and Stilgoe, J. *Why Responsible Innovation?,* in Responsible Innovation: Managing the Responsible Emergence of Science and Innovation in Society, edited by R. Owen, J. R. Bessant, and M. Heintz, 306, OXFORD: WILEY (2013).

[30] *See* U. Gasser and V. Almeida, *A layered model for AI governance*, IEEE INTERNET COMPUTING, vol. 21, no. 6, pp. 58-62, November/December 2017, doi: 10.1109/MIC.2017.4180835.

[31] *See e.g.,* C. Emerson, S. James, K. Littler, and F. (Fil) Randazzo, *Principles for gene drive research*, SCIENCE 358, ISSUE 6367, PP. 1135-1136, 1 DEC (2017) DOI: 10.1126/science.aap9026

[32] *See e.g.,* Hursthouse, Rosalind and Glen Pettigrove, *Virtue Ethics*, THE STANFORD ENCYCLOPEDIA OF PHILOSOPHY (Winter 2022 Edition), Edward N. Zalta & Uri Nodelman (eds.), https://plato.stanford.edu/archives/win2022/entries/ethics-virtue/

*Third,* we do not intend the framework we propose to be translated into binding law, at least not initially. Its effectiveness will therefore consist in voluntary commitments, at first. For those who seek to practice RRI in the context of QT, it can serve as a useful point of departure. In time, a framework like this could become an effective basis for self-regulation or evolve into instruments of soft and – eventually – hard law, such as a Quantum Governance Act.

As shown in Figure 1, such combined quantum-specific and general principles could be categorized into an overarching framework with distinct functional categories that balance the need to support, protect, and incentivize QT *advancements* with the need to establish appropriate *safeguards,* while *engaging* society. In other words, the technical, ethical, and social/legal/policy levels can be visualized as layers on which the principles that guide responsible quantum innovation can be formulated per SEA category.

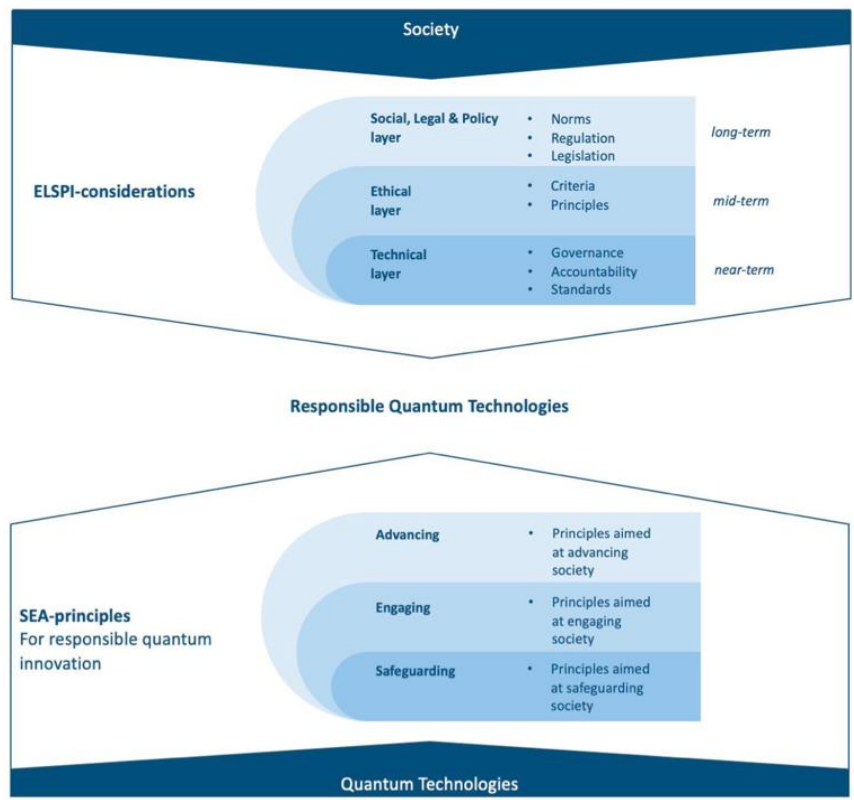

*Figure 1. SEA-framework for Responsible Quantum Technology*[34]

---

[33] The envisioned Principles for Responsible Quantum Technology would likely benefit from general principles that also apply to other technologies, such as AI, nano, biotech, and nuclear, *see e.g.,* M. Kop & M. Brongersma, *Integrating Bespoke IP Regimes for Quantum Technology into National Security Policy*, August 8, 2021, Stanford Law School, Working Paper, https://law.stanford.edu/publications/integrating-bespoke-ip-regimes-for-quantum-technology-into-national-security-policy/; and Gasser & Almeida, *supra* note 30.

[34] This methodological framework builds on the work of Gasser and Almeida on the layered model for technology governance, *see* Gasser & Almeida, *supra* note 30.

(1) *Safeguarding*-principles should address the downside risks arising from QT (Fig. 1). They aim to protect society by taking appropriate measures.[35] For instance, two safeguarding principles could be: *i)* consider information security as an integral part of QT R&D and *ii)* proactively anticipate the potential malicious dual use of QT applications. Operationalizing these principles would promote approaches such as implementing risk-based quantum impact assessments -overseen by product and program managers- to minimize security threats, performing risk-reward analysis before granting production and market authorization, or reducing the risks that QT can be used for harmful purposes. Examples of these so-called *dual uses* include (1) the use of quantum simulation to help develop new drugs, fertilizers or industrial catalysts versus the manufacturing of new biological or chemical weapons using the same foundational technology involving scalable quantum hardware, and (2) using photonics and plasmonics to miniaturize quantum devices to improve energy efficiency - benefitting the planet, versus utilizing miniaturization in quantum sensors to watch people's every move - infringing core human rights.[36]

(2) *Engaging*-principles should enhance societal engagement of QT innovators and other stakeholders (Fig. 1). They should address issues such as the threat of a deepening quantum gap among countries, the need for adequate intellectual property (IP) and fair competition mechanisms for QT, and pursue building diverse quantum communities.[37] For example, IP in this context should be calibrated in a manner that balances incentives, rewards, access, and risks, while addressing associated geostrategic and national security concerns. Moreover, IP policies should work in concert with tailored antitrust regulations to prevent unwanted winner-takes-all market behavior within the emerging quantum ecosystem.[38]

(3) *Advancing*-principles should progress society through QT-based innovation (Fig. 1). Advancing principles should both promote QT R&D -creating a virtuous cycle of progress that fueled Moore's law similar to that of the semiconductor industry- and encourage QT applications for desirable social goals.[39] The 17 UN's Sustainable Development Goals[40] - which are a call for action by both developed and developing countries to adopt a global 2030 agenda for peace and prosperity - might serve as an illustration for objectives to which QT can contribute to areas such as drug discovery, resource optimization, water management, 21-day weather forecasting to improve agriculture, and climate modeling. Additionally, QT could help to enhance complementary innovation including quantum-classical synergies such as the variational quantum eigensolver (VQE), and quantum-AI hybrids such as quantum machine learning and specialized quantum co-processors for AI that solve optimization problems that are typically hard for classical systems.[41]

---

[35] *See* for a call to action to address risks arising from QT, Khan, I, *Will Quantum Computers Truly Serve Humanity?* SCIENTIFIC AMERICAN, 2021, https://www.scientificamerican.com/article/will-quantum-computers-truly-serve-humanity/.

[36] For further reading on quantum technologies becoming instrumental in both security and warfare, *see* Zhou Q., *The subatomic arms race: Mutually assured development,* HARVARD INTERNATIONAL REVIEW, (2021), https://hir.harvard.edu/the-subatomic-arms-race-mutually-assured-development/.

[37] *See* Aboy, Minssen & Kop, *supra* note 6. Compare to M. Kop, *Quantum Computing and Intellectual Property Law,* 25 BERKELEY TECHNOLOGY LAW JOURNAL 2021, PP 101-115, https://btlj.org/2022/02/quantum-computing-and-intellectual-property-law/. For an unorthodox yet creative take on empirical IP research in the field of QT, *see* Zeki Can Seskir & Kelvin W. Willoughby, *Global Innovation and Competition in Quantum Technology, Viewed Through the Lens of Patents and Artificial Intelligence*, INTERNATIONAL JOURNAL OF INTELLECTUAL PROPERTY MANAGEMENT, 13, 1 (2023), 40-61, https://dx.doi.org/10.1504/IJIPM.2021.10044326

[38] *See* Kop, Aboy & Minssen, *Intellectual property in quantum computing and market power: a theoretical discussion and empirical analysis*, JOURNAL OF INTELLECTUAL PROPERTY LAW & PRACTICE, VOLUME 17, ISSUE 8, Oxford University Press, August 2022, Pages 613–628, https://doi.org/10.1093/jiplp/jpac060

[39] *See* National Academies of Sciences, Engineering, and Medicine. 2019. *Quantum Computing: Progress and Prospects*. Edited by Mark Horowitz and Emily Grumbling, Washington, DC: The National Academies Press, pp158-159. https://doi.org/10.17226/25196. A virtuous cycle, or positive feedback loop, requires continuous private and public funding, research, development and engineering efforts, attracting talent, successful commercial QT applications, increasing demand for QT, and economies of scale driving progress, growth and profitability.

[40] *See* https://sdgs.un.org/goals

*Balancing safeguarding, engaging, and advancing (SEA)*

In general, engaging principles should help guide and promote a fit-for-purpose QT regulatory framework that balances the *safeguarding* and *advancing* principles to promote quantum innovation. Engagement principles should encourage inclusivity and competition. As such, they interact with safeguarding and advancing principles. For instance, engaging principles should help manage the risks of a few dominant QT players achieving monopolistic competitive advantage by restricting access to essential QT infrastructure (*safeguarding*). That said, the engagement principles should also interact with *advancing* principles to help promote access to key QT infrastructures by encouraging open access to cloud-based quantum computers, open quantum interoperability standards and protocols, and open-source quantum development tools.[42] Such initiatives lower the education barriers of entry to build a skilled quantum workforce by empowering those with an internet connection to start working with quantum computers at the level of pulses, gates, circuits, and application modules. The ability to program and prototype quantum algorithms using Python, open quantum assembly language (QASM), or utilizing a graphical circuit composer to implement quantum algorithms on a real quantum computer accessible through the cloud, substantially increases engagement and promotes inclusivity. Here, access encourages creativity, experimentation and simply finding out what works and what does not. Finally, engaging principles should encourage awareness of relevant QT issues in society through general education and cross-sector dialog with stakeholders across all layers of society, including QT developers, investors, regulators, and the public.[43]

Articulating the foundational principles for Responsible QT clustered along the proposed three functional dimensions could be inspired by established best practices of responsible research and innovation (RRI).[44] The proposed framework would need to be operationalized by guiding principles that should be incorporated into the design, architecture, and infrastructure of QT systems, products, and services on a global scale, resulting in Responsible QT *by design and default*.[45] Crucially, this requires translating the principles into specific R&D design decisions, continuously testing, benchmarking, and validating results, verifying their usefulness, adjusting where deemed appropriate. Concretely, designing quantum computing and sensing hardware architectures, creating quantum software as a service (SaaS) platforms, formulating quantum algorithms, and building a future Quantum Internet must each adhere to the principles of Responsible QT and their underlying norms, standards, and values, aiming for a responsible quantum ecosystem and fostering sustainable innovation.

---

[41] This includes quantum/AI hybrids and using quantum resources in AI such as quantum assisted machine learning, hybrid cloud computing, quantum-classical interfaces, and quantum/AI simulation on classical systems. As quantum computing could be a major boost for AI it is necessary to proactively address present day ethical problems pertaining to AI such as bias, representativeness, black boxes, and polluted data, in order to ensure that quantum computing doesn't exacerbate these problems. Here too, our proposed framework can play an important anticipatory role, complementing existing frameworks for AI. That said, these AI issues are primarily related to the *data* aspects (e.g., the coverage and generality of the training data) as opposed to the *computational* aspects.

[42] For further reading on quantum technology and standardization, *see* DeNardis, Laura, *Quantum Internet Protocols* (August 4, 2022), http://dx.doi.org/10.2139/ssrn.4182865

[43] *See e.g.,* Potomac Quantum Innovation Centre, *Our Quantum Future: Some Assembly Required,* (2022), https://www.quantumworldcongress.com/whitepaper

[44] For an overview of RRI literature, *see* R. Kumar Thapa, T. Iakovleva, L. Foss *Responsible research and innovation: a systematic review of the literature and its applications to regional studies*, EUROPEAN PLANNING STUDIES, 27:12, 2470-2490, (2019). DOI: 10.1080/09654313.2019.1625871 *See* in similar vein, Coenen, C., Grinbaum, A., Grunwald, A., Milburn, C., & Vermaas, P. *Quantum Technologies and Society: Towards a Different Spin,* NANOETHICS, 16, 1–6 (2022), https://doi.org/10.1007/s11569-021-00409-4

[45] For ethical design thinking in technology, *see* B. Friedman, *Embodying values in technology: Theory and practice*, INFORMATION TECHNOLOGY AND MORAL PHILOSOPHY, 3(6):322–353, CAMBRIDGE PUBLISHER: CAMBRIDGE UNIVERSITY PRESS, (2008).

Connecting both RRI dimensions and ELSPI considerations to responsible quantum R&D, categorized along the lines of the SEA-framework's trifecta of Safeguarding, Engaging, and Advancing QT, this catalogue of principles for Responsible QT could be imagined as follows:[46]

1. *Consider information security as an integral part of QT, addressing security threats;*
2. *Proactively anticipate the malicious use of quantum applications, addressing risks of dual use;*
3. *Seek international collaboration based on shared values, addressing a winner-takes-all dynamic;*
4. *Consider our planet as the sociotechnical environment in which QT should function, engaging states;*
5. *Be as open as possible, and as closed as necessary, engaging institutions;*
6. *Pursue diverse quantum R&D communities in terms of disciplines and people, engaging people;*
7. *Link quantum R&D explicitly to desirable social goals, advancing society;*
8. *Actively stimulate sustainable, cross-disciplinary innovation, advancing technology;*
9. *Create an ecosystem to learn about the possible uses and consequences of QT applications, advancing our understanding of Responsible QT;*
10. *Facilitate dialogues with stakeholders to better envision possible quantum futures, advancing our collective thinking and education about QT and its impact.*

As future QT scenarios remain largely unknown, epistemic modesty must be embedded throughout this process. Put differently, as the field is still in its infancy, the framework might have to be adapted to new discoveries. Operationalizing these guiding principles would therefore call for continuous collaborative multi-stakeholder and industry efforts that follow and steward the life cycle of QT systems, products, and services and anticipate novel quantum use cases,[47] for instance involving standard-setting organizations such as ISO and NIST, as well as professional organizations such as the IEEE. Research institutions and research-driven companies would need to dedicate resources in the form of IRB-like bodies (institutional review boards), assessing the ethical implications of a particular QT.[48] Ideally, such the results of such assessments and oversight mechanisms are openly shared to advance a collective learning process and inform evidence-based policy making.

Training and continued education programs should help to further develop and implement the principles and drive QT innovation across economic and industrial sectors, such as biopharma, energy, mining, communications, logistics, defense, and space. Guided by the proposed set of principles, these networked communities of practice would identify and address ELSPI challenges in their respective R&D and application contexts, including dual uses of QT, IP, fair competition, product safety and liability, inclusion, complementary innovation, equitable distribution of QT's benefits, innovation externalities, spillovers and trade-offs, capability overhang, and the global quantum race.[49] The importance of such practices can be demonstrated by having a closer look at the example of information security in the post-quantum era.

---

[46] For a full description of the catalog of Principles for Responsible Quantum Innovation within the context of key identified topics and aims, including detailed illustrations per principle, s*ee* Kop *et al., 10 Principles for Responsible Quantum Innovation*, forthcoming 2023.

[47] Pivotal work must be done to discover novel quantum use cases beyond cybersecurity and optimize risk-benefit curves before their inevitable dual use characteristics can harm society, *see in similar vein* Johnson, W. G., *Governance tools for the second quantum revolution,* Jurimetrics, 59, 487 (2018), https://www.jstor.org/stable/27009999

[48] Originally, an IRB focuses on rights and welfare -as in quality of life- of human subjects in research. In the case of QT, especially at the current early stage, it would be relevant to expand this scope to allow for broader QT impact assessments. For institutionalizing RRI in this context, see Owen, R., M. Pansera, P. Macnaghten, and S. Randles, *Organisational Institutionalisation of Responsible Innovation*, RESEARCH POLICY 50 (1): 104132 (2021), doi:10.1016/j.respol.2020.104132.

## V. ILLUSTRATING THE CHALLENGE: INFORMATION SECURITY IN THE POST-QUANTUM ERA

Quantum algorithms have the potential to break current cryptography protocols,[50] threatening the information security of existing information technologies (IT) and the privacy of its users.[51] This could destabilize society and undermine trust in its institutions. QT could expose extensive swaths of information currently regarded as private and confidential, ranging from sensitive personal data to financial sector and national security information assets.[52] Concretely, we already have quantum algorithms capable of breaking our widespread public key cryptosystems as soon as the quantum computer hardware is sufficiently mature. Thus, we can foresee potentially disruptive effects to fundamental human rights such as privacy and data protection[53], including large scale loss of privacy, human identity theft, loss of confidentiality and integrity of digital communication on the Internet, obstruction of commercial transactions, leaking of highly sensitive trade and state secrets, and other unwanted global surveillance disclosures.[54]

Given that information security threats are among the most pressing themes in the context of Quantum-ELSPI, one of the *safeguarding* principles should promote information security to be a central feature of Responsible QT. That said, the challenge is to achieve appropriate *safeguarding* while continuing to promote *advancement* of QT such as quantum computing (QC). After all, QC holds great promises for societal benefits because quantum algorithms can solve hard computational problems that are mathematically intractable for classical computers. However, as explained above in section 2, these same problems that are hard for classical computers to solve have been selected as a fundamental building block of our widespread public key cryptosystems precisely because they were believed to be computationally intractable.

The possibility of implementing a "store now, decrypt later" strategy should provide incentives to start replacing our existing cryptosystems for critical information assets with quantum-safe cryptographic systems that are resistant to attacks by quantum algorithms as early as possible. It could take years to replace our cryptosystems. In fact, it has taken nearly two decades to deploy our modern asymmetric key cryptographic infrastructure. Thus, once sufficiently fault-tolerant quantum computer hardware is available, it could be used to reveal information assets previously encrypted with our most common forms of public key cryptography. This calls for researching and investing in quantum-safe cryptography initiatives, as well as *advancing and engaging in* quantum-safe information security programs.[55]

---

[49] *See* M. Kop, *Establishing a Legal-Ethical Framework for QT,* YALE J.L. & TECH. THE RECORD (2021). https://yjolt.org/blog/establishing-legal-ethical-framework-quantum-technology
[50] *See e.g.,* Aboy, Minssen & Kop, *supra* note 6.
[51] FACT SHEET: *President Biden Announces Two Presidential Directives Advancing QT*, White House (May 4, 2021). https://www.whitehouse.gov/briefing-room/statements-releases/2022/05/04/fact-sheet-president-biden-announces-two-presidential-directives-advancing-quantum-technologies/
[52] *Id. See also* Roberson, T., Leach, J., & Raman, S. *Talking about public good for the second quantum revolution: Analysing quantum technology narratives in the context of national strategies*, QUANTUM SCIENCE AND TECHNOLOGY, 6(2), 25001, (2021), https://doi.org/10.1088/2058-9565/abc5ab; and Hoofnagle & Garfinkle, *supra* note 7.
[53] *See also* M. Wimmer, T. Moraes, *Quantum Computing, Digital Constitutionalism, and the Right to Encryption: Perspectives from Brazil*. DISO 1, 12, SPRINGER NATURE (2022). https://doi.org/10.1007/s44206-022-00012-4
[54] Kop, *supra* note 49.

From a *safeguarding* standpoint we must ensure that the field of post-quantum cryptography is on par, or ahead (*advancing*) of the realization of a fault-tolerant or error-corrected quantum supercomputer with enough qubits to implement efficient decryption algorithms that leverage quantum parallelism to break our widespread cryptosystems. In short, large scale physical realizations of quantum computers capable of deploying Shor's algorithm to break RSA-2,408 should be paralleled by the development of quantum-resistant cryptographic algorithms. Such algorithms must be safe and secure against cryptanalytic attacks by QC that employ quantum algorithms breaking common asymmetric (public) key cryptographic systems in use today.[56] This is accomplished by exploiting mathematical problems that are computationally intractable, i.e. for which we do not have a known efficient solution using classical or quantum computers. NIST has initiated a broad *engaging* process to solicit, evaluate, and standardize quantum-resistant public-key cryptographic algorithms.[57] The goal is to create cryptographic systems that are secure against both classical computers and quantum computers.[58] Additionally, these cryptographic systems should be able to interoperate with existing protocols and networks.

From an *advancing* standpoint, it is noteworthy that QT developments can also help to improve security and data privacy in the sense of safe transfer of information.[59] Some use cases already promise to benefit from quantum cryptography advances such as quantum key distribution (QKD) to achieve secure communications by implementing quantum-based cryptographic protocols to exchange symmetric keys. QKD uses fundamental quantum-mechanical properties to securely exchange keys over an unsecure public channel. Additionally, several protocols have recently emerged to enable private quantum computation. These protocols, such as blind quantum computation (BQC) are designed to secure computation rather than communication.[60]

At the operational level, research on and development of QT should be accompanied by risk-based quantum impact assessments focused on information security risks and implementing controls to mitigate such risks. This includes the implementation of state-of-the-art information security management systems (ISMS) such as ISO27001 to protect information assets from a particular QT R&D program. Notably, this requires extending the ISO 27001/27002 controls to include the implementation of quantum-safe information security controls.

---

[55] *See* NIST Post-Quantum Cryptography https://csrc.nist.gov/projects/post-quantum-cryptography
[56] *See* Shor, *supra* note 13.
[57] *See* NIST Announces First Four Quantum-Resistant Cryptographic Algorithms, NIST, July 05, 2022, https://www.nist.gov/news-events/news/2022/07/nist-announces-first-four-quantum-resistant-cryptographic-algorithms
[58] *See, e.g.,* J. Yin, Y. Li, S. Liao *et al. Entanglement-based secure quantum cryptography over 1,120 kilometres.* NATURE 582, 501–505 (2020). https://doi.org/10.1038/s41586-020-2401-y
[59] For an architecture combining Quantum Random Access Memory (QRAM) and quantum networks, resulting in multi-party private quantum communication, *see.*Junyu Liu, Connor T. Hann, Liang Jiang, *Quantum Data Center: Theories and Applications*, August 1, 2022, https://arxiv.org/abs/2207.14336.
[60] *See, e.g.,* J. Fitzsimons, *Private quantum computation: an introduction to blind quantum computing and related protocols.* NPJ QUANTUM INF 3, 23 (2017). https://doi.org/10.1038/s41534-017-0025-3

Page 14 of 22

# VI. CHARTING THE PATHWAY FORWARD: EMERGING REGULATION OF QUANTUM TECHNOLOGY

This section briefly examines how our proposed SEA-framework for Responsible QT can inform the emergent regulatory landscape affecting QT, taking the two recent Executive Orders signed May 4, 2022, by President Biden, and the Quantum Computing Cybersecurity Preparedness Act that became public law on December 21 2022, as examples.

The framework can contribute in three analytically distinct, yet practically intertwined ways. *First,* it offers an *analytical* lens to help examine the role, purpose, and process dimensions of emerging regulations, whether general in nature or specifically tailored to QT, to identify and examine such issues in a respective application context and regulatory field. As noted, along recent QT advancements comes the need for careful consideration of legal, regulatory, ethical, and policy issues to ensure that these technologies are safe and ethically sound.[61] This is an area where the SEA-framework can offer not only substantive anchors, but also procedural guide shared among various QT stakeholders to unlock the potential of 'regulation as facilitation'.[62] Regulation of novel technologies should be flexible and dynamic to foster innovation while still providing adequate safeguards to protect against potential harms. Both our dimension of *engagement* and the concept of 'Regulation as facilitation' emphasize collaboration between regulators, industry, and other stakeholders. They help ensure technology is developed in a way that maximizes their potential benefits while minimizing risks and avoiding unnecessary regulatory burdens.

*Second*, the SEA-framework can serve as a reference point to *evaluate or assess* regulatory initiatives regarding the critical question as to what extent such regulatory endeavours strike a balance between safeguarding, engaging, and advancing QT. For instance, it might reveal to what extent proposed future QT regulation in the EU might follow the approach taken in AI by putting strong emphasis on safeguarding QT fueled by the precautionary principle, compared to the US permissionless innovation approach, which is expected to emphasize advancing to retain global tech leadership[63] by actively pursuing a democratic values-based quantum ecosystem in a US-led liberal global order.[64]

Here, the frameworks' *third* contribution becomes visible: It offers at least some initial *guideposts* of how regulatory interventions for quantum as a base-layer/general-purpose technology could be designed, contextualized, and tailored to their exceptional nature, balancing open innovation and risk control.[65] Specifically, the proposed SEA-framework offers clues on how to balance safeguarding and advancing QT, aiming for Responsible QT by design and default. Similarly, regulating QT can be thought of as a balancing act between under-regulation and over-regulation.[66] The framework can be helpful to raise awareness of important issues in this context. The success of this joint optimization will also depend on the strength of the framework's *engaging* component which emphasizes the importance of ongoing dialogue between regulators, experts, and industry to ensure regulation remains effective and responsive to emerging technological developments. Without such engagement it is likely that the regulatory focus would seek to primarily optimize one of these dimensions (e.g. safeguarding). The issue with such approach is that optimizing for QT safeguarding would likely result in less safety, since adequate safeguarding is dependent on further QT advances, for instance, in quantum cryptography.

---

[61] For further reading on the potential impact of the use of QC in the legal sector, *see* Jeffery Atik & Valentin Jeutner, *Quantum computing and computational law*, LAW, INNOVATION AND TECHNOLOGY, 13:2, 302-324, 2021, DOI: 10.1080/17579961.2021.1977216

[62] Black, Julia, *Regulation as Facilitation: Negotiating the Genetic Revolution*, THE MODERN LAW REVIEW 61, NO. 5 (1998): 621–60. http://www.jstor.org/stable/1097126

[63] Multidisciplinary embedded Responsible QT has significant competitive advantages, fostering exponential innovation and trust, while steering toward beneficial societal outcomes.

[64] Although the as of yet mostly unregulated quantum sphere offers a once in a lifetime opportunity to harmonize legal-ethical frameworks for QT on the planetary level, we anticipate global divergence in QT regulation between the three tech blocks US, EU, and China, who each have their own values systems, which are culturally sensitive, context-specific, and dynamic. Whoever sets the technical standards will set the rules of the road for quantum for the world to follow.

[65] *Compare to* Yang, Jialei & Chesbrough, Henry & Hurmelinna, Pia, *How to Appropriate Value from General-Purpose Technology by Applying Open Innovation*, CALIFORNIA MANAGEMENT REVIEW 2021, 000812562110417, 10.1177/00081256211041787, at 18-20.

Taken together, the suggested SEA-framework for Responsible QT and its operationalizing principles might inform current and future regulations that share the challenge of optimizing among the different dimensions and issues detailed in this article. Creating horizontal (federal, or international level) norms for QT as a base-layer technology[67] that apply across industrial and economic sectors is a pressing tasks that should be taken on before the technology becomes locked in.[68] But it is also a challenging task, as any regulation must consider the exceptional, counter-intuitive traits of applied quantum physical phenomena, its unseen functionality, and their potential for dual use[69] - balancing open innovation, value appropriation, IP, fair trade and competition, and risk control.[70] In addition, the inherent uncertainty underpinning emerging QT[71] applied to real world systems, products, and use cases (e.g., quantum simulation of molecular physics and biochemistry across a wide range of dual-use applications), calls for a risk based approach that incentivizes sustainable innovation in parallel, e.g. via regulatory sandboxes that afford breathing room for experimentation and prototyping.[72] To the extent the proposed SEA-framework for Responsible QT is informed by the technical and physical underpinnings of QT, it might in turn help to ensure that emerging rules and codified laws for QT live up to the same challenge.

---

[66] An example of under-regulation would be to have the market figure out how to self-regulate laissez-faire style, potentially only benefitting a small select group of corporations instead of society at large.

[67] Several quantum technologies such as "quantum computers" should be considered "base-layer technologies" akin to a microprocessor (CPU). In the same way that it did not make sense to focus the regulations on the classical microprocessor (the fundamental building block of classical computation) but instead direct the regulation to the upstream devices and applications using these microprocessors, the primary regulatory efforts -with a few exceptions such as of export control- should not be directed to "quantum processors" or "quantum sensors" but instead target the upstream QT-enabled solutions for specific applications using the legal frameworks available for such domains. For instance, a medical device using a microprocessor is regulated as a medical device due to its intended use satisfying the FDA or EU MDR regulatory definition of medical device while a gaming console using the same microprocessor does not follow these regulations. In the same way, if a medical device were to make use of cloud-based access to a quantum computer it would be regulated as a medical device without raising any new legal issues (i.e., the nature of the processing element is abstracted from the regulation which focuses on the intended use), while access to the same cloud-based quantum computer for education-purposes (e.g., running basic quantum algorithms as part of a university course in quantum computing) or research (e.g., quantum simulation of molecules) will not. Similarly, graphical processing units (GPUs) are the computational building blocks enabling AI, but also for gaming. Thus, care must be exercised to ensure that regulatory interventions target the correct layer in the stack.

[68] *See* M. Kop, *Regulating Transformative Technology in The Quantum Age: Intellectual Property, Standardization & Sustainable Innovation* (October 7, 2020), STANFORD - VIENNA TRANSATLANTIC TECHNOLOGY LAW FORUM, TRANSATLANTIC ANTITRUST AND IPR DEVELOPMENTS, STANFORD UNIVERSITY, ISSUE NO. 2/2020, https://law.stanford.edu/publications/regulating-transformative-technology-in-the-quantum-age-intellectual-property-standardization-sustainable-innovation/ *Compare to* Johnson, *supra* note 47.

[69] *See e.g.,* E. Perrier, *The Quantum Governance Stack: Models of Governance for Quantum Information Technologies*, DIGITAL SOCIETY, QUANTUM-ELSPI TC, 1, 22, SPRINGER NATURE, (2022).

[70] *See also* Kop, *supra* note 49.

[71] For further reading on the uncertainty principle in the quantum physics underlying QT, *see* Feynman et. al., *supra* note 14.

[72] *See* Kop, *supra* note 68. *Compare with* Dekker, T. and Martin-Bariteau, F., *Regulating Uncertain States: A Risk-Based Policy Agenda for Quantum Technologies* (May 1, 2022). (2022) 20:2 CANADIAN JOURNAL OF LAW AND TECHNOLOGY 179, OTTAWA FACULTY OF LAW WORKING PAPER NO. 2022-26, http://dx.doi.org/10.2139/ssrn.4203758

***Regulatory Approaches to the Emerging Quantum Technology Landscape***

The previous section has described in the abstract how the SEA-framework for Responsible QT can inform emerging efforts aimed at regulating QT by offering analytical, evaluative, and design baselines. Recent developments in the US that might mark some of the cornerstones of the future regulatory landscape can serve as a use case to illustrate how the Responsible QT framework might be used in practice when examining emerging QT legal and regulatory norms and approaches.

Recent building blocks of what might be an emerging legal framework include two Executive Orders signed on May 4 2022 by President Biden. The directives are aimed at *advancing* US quantum information science *(QIS)* by laying "the groundwork for continued American leadership in an enormously promising field of science and technology […] to foster these advances by furthering the President's commitment to promoting breakthroughs in cutting-edge science and technology," *safeguarding* "while mitigating the risks that quantum computers pose to America's national and economic security […]", and *engaging* "it does so by enhancing the National Quantum Initiative Advisory Committee, the Federal Government's principal independent expert advisory body for quantum information science and technology."[73] The *safeguarding* strategy includes both elements to address the direct proximate risks "plan to address the risks posed by quantum computers to America's cybersecurity", as well as broader risk-based considerations surrounding American IP by urging "Federal agencies to develop comprehensive plans to safeguard American intellectual property, research and development, and other sensitive technology from acquisition by America's adversaries, and to educate industry and academia on the threats they face". Thus, the directives effectively connect IP protection to national and economic security strategy, while emphasizing the importance of *advancing* responsible and secure R&D in quantum computing.[74] In addition, the *Quantum Computing Cybersecurity Preparedness Act of December 2022* requires The Office of Management and Budget by law to give priority to federal agencies' purchases of and transitions to post-quantum cryptographic IT systems, with the goal of *safeguarding* through *advancement* of post-quantum cryptography.[75] Together with the bipartisan supported America Competes Act and the CHIPS and Science Act[76][77] which are applicable to semiconductors including those used in QT, these Acts aim to strengthen supply chains, and counter US adversaries such as China.[78] Collectively, these directives intend to lay the groundwork for continued American competitiveness and leadership in QT, fostering innovation while mitigating risks associated with dual use quantum technology by instigating targeted export controls.[79]

---

[73] FACT SHEET: President Biden Announces Two Presidential Directives Advancing Quantum Technologies, White House (May 4, 2021), https://www.whitehouse.gov/briefing-room/statements-releases/2022/05/04/fact-sheet-president-biden-announces-two-presidential-directives-advancing-quantum-technologies/. *See also* https://www.quantum.gov/
[74] Washington is clearly taking the lead to create the rules of the road for quantum for the rest of the world to follow, effectively protecting US intellectual property from theft by global competitors such as China and Russia, which are at present systemic rivals [with incompatible ideologies]. While Brussels is focusing on AI safeguarding, Washington attention is turning to ensuring technological leadership in the post-quantum era.
[75] *See* https://www.congress.gov/bill/117th-congress/house-bill/7535/actions, and https://fedscoop.com/biden-signs-quantum-computing-cybersecurity-act-into-law/
[76] *See* CHIPS & Science Act of 2022, https://www.commerce.senate.gov/services/files/2699CE4B-51A5-4082-9CED-4B6CD912BBC8
[77] *See, e.g.,* McKenzie Prillaman, *Billions more for US science: how the landmark spending plan will boost research*, NATURE 608, 249 (2022) https://doi.org/10.1038/d41586-022-02086-z
[78] *See* https://www.whitehouse.gov/briefing-room/statements-releases/2022/08/09/fact-sheet-chips-and-science-act-will-lower-costs-create-jobs-strengthen-supply-chains-and-counter-china/ and https://dean.house.gov/2022/2/the-house-passes-the-america-competes-act
[79] *See in this context also* Implementation of Certain New Controls on Emerging Technologies Agreed at Wassenaar Arrangement 2018 Plenary, Federal Register 2019,

When analyzing and assessing these regulations through the looking glass of the proposed SEA-framework for Responsible QT, various components crystallize that map onto three SEA categories. For example, catalyzing investments in domestic advanced semiconductor manufacturing capacity and instigating export controls can be considered an attempt to jointly optimize both *safeguarding* and *advancing* US competitiveness in quantum computing. Similarly, the emphasis in investing in post-quantum cryptography illustrates how *safeguarding* can be achieved by *advancing*, as opposed to attempting to achieve safeguarding objectives by limiting technological development. Connecting intellectual property strategy to national security policies also illustrates the overarching aim of jointly optimizing the advancing, safeguarding and engagement dimensions.[80] These regulatory objectives to map our suggested catalogue of 10 principles for Responsible QT listed above in section IV, e.g. Principle 5: *'Be as open as possible, and as closed as necessary'*.[81]

Further, section 1 of the Executive Order on Enhancing the National Quantum Initiative Advisory Committee aims to enable knowledge transfer between industry, academia and government to ensure American leadership in QT, especially quantum information technologies.[82] This relates to the SEA categories of engaging and advancing. The goal is to achieve QT leadership (advancement) through engagement. Section 1 b of the *National Security Memorandum on Promoting United States Leadership in Quantum Computing While Mitigating Risks to Vulnerable Cryptographic Systems* focuses on the significant risks QC potentially poses to the economic and national security of the United States, which ties to the SEA category of safeguarding.[83] The purpose of the Quantum Computing Cybersecurity Preparedness Act is '*to encourage the migration of Federal Government information technology systems to quantum-resistant cryptography, and for other purposes.*'[84] While this mainly qualifies as safeguarding society as described in our case study of information security in the post-quantum era, it aims to achieve this safeguarding primarily through *advancement* in the field of post-quantum cryptography. This includes R&D in quantum-based solutions (i.e. protecting against quantum-crypto attacks using quantum technologies), as well as the search for cryptosystems based on problems that are computationally intractable to both classical and quantum computers. Thus, the goal should be to achieve the safeguarding objectives by further advancing QT in the sense that novel algorithms, software, and hardware solutions will have to be developed, increasing economic growth and competitiveness. In addition, the Act offers a chance to build a multidisciplinary, intergenerational quantum workforce as suggested by Principle 6 above, stating to: *'Pursue diverse quantum R&D communities in terms of disciplines and people, engaging people'*. Notably, quantum computing

---

https://www.federalregister.gov/documents/2019/05/23/2019-10778/implementation-of-certain-new-controls-on-emerging-technologies-agreed-at-wassenaar-arrangement-2018

[80] For a podcast discussion about QT's legal and ethical implications on law, economics, and society with a focus on US attempts to incorporate the technology into existing frameworks for intellectual property and international security, *see* Quantum Computing with Joonas Keski-Rahkonen and Katri Nousiainen, August 23, 2022, Berkeley Technology Law Journal Podcast: Quantum Computing, https://btlj.org/2022/08/berkeley-technology-law-journal-podcast-quantum-computing/

[81] *See also* Kop et al., *supra* note 46.

[82] *See* Executive Order on Enhancing the National Quantum Initiative Advisory Committee, https://www.whitehouse.gov/briefing-room/presidential-actions/2022/05/04/executive-order-on-enhancing-the-national-quantum-initiative-advisory-committee/

[83] *See* National Security Memorandum on Promoting United States Leadership in Quantum Computing While Mitigating Risks to Vulnerable Cryptographic Systems, https://www.whitehouse.gov/briefing-room/statements-releases/2022/05/04/national-security-memorandum-on-promoting-united-states-leadership-in-quantum-computing-while-mitigating-risks-to-vulnerable-cryptographic-systems/

[84] *See* Quantum Computing Cybersecurity Preparedness Act, https://www.congress.gov/bill/117th-congress/house-bill/7535/text

brings together the more established American workforces in quantum physics, chip design, and semiconductor manufacturing that has led in global competitiveness from 1950 to the 2000s, with the innovative software workforce that has led the major high-tech developments over the last 20 years.

When moving from the analytical and evaluative dimensions of the SEA-framework to the prospective *question of design*, additional considerations might come into play. Given the current lack of comprehensive best practices in terms of Responsible QT-oriented policymaking, and considering the need for policymakers to make decisions anyway and under conditions of uncertainty, the framework's normative power to provide specific substantive guidance is limited. However, as discussed in Section IV, the framework alludes to a broad range of options and approaches that policymakers can embrace to steer the development of Responsible QT environments, which offers at least minimum guidance in terms of "asking the right questions" considering the full options available and probing their respective SEA implications. The framework also helps avoid potential detrimental policies-mistakes such as merely optimizing a single dimension (e.g. trying to achieve safeguarding objectives through unbalanced regulatory efforts that hinder further technological development instead of promoting and leveraging QT advancements in order to achieve these underlying safeguarding goals). Further, the framework highlights the important role *self-regulation* might play across technical, ethical, and societal levels by prescribing pro-active measures such as open-source quantum software and hardware movements, quantum impact assessments, and technological measures to safeguard human rights and freedoms.

Moreover, the SEA-framework indicates that such self-regulatory approaches might work in concert with *"hard law"*, for instance in gestalt of an emerging legal and infrastructural ecosystem consisting of overarching horizontal rules for a particular general purpose technology – analogous to the EU AI Act – flanked by industry-specific legal frameworks. This could cumulate in a binding (International) Quantum Governance Act, or a Global Quantum Treaty[85], to be enforced by hyperspecialized overseeing bodies such as the FDA or designated notified bodies. Such an extensive set of QT standards would have to interoperate with other areas of the legal system and embedded in existing regulatory structures.[86] This raises some tensions, as we find ourselves in a continuum moving from classical to quantum, where interwoven physical characteristics and legal designations are a matter of degree, and cannot be clearly separated from each other. For example, the (diffuse) technical classification of quantum information in the cloud as data in the classical sense, and defining a quantum/AI hybrid chip as both QT and AI device has consequences for applicable legal regimes.[87] Moreover, these future legal regimes pertaining to quantum information should ideally be applicable to across the range of QT, including quantum computing, simulation, sensing, and communication.[88]

---

[85] In a similar vein, the Council of Europe's CAI - Committee on Artificial Intelligence is currently drafting a Convention on Artificial Intelligence, Human Rights, Democracy and the Rule of Law, *see* https://rm.coe.int/cai-2023-01-revised-zero-draft-framework-convention-public/1680aa193f This is a different initiative than the imminent EU AI Act as proposed by the European Commission, *see* https://digital-strategy.ec.europa.eu/en/policies/european-approach-artificial-intelligence

[86] More research is required to ascertain whether these rules should have the form of a Presidential executive order, or in the form of a more durable new law from Congress, whether it would work for the tech industry, and whether it's scope should be broad or narrow.

[87] *See* note 67 about microprocessors, GPUs, and quantum computing as "base-layer technologies". Efforts to classify hybrid quantum-classical computing paradigms are underway, *see* Frank Phillipson, Niels Neumann, Robert Wezeman, *Classification of Hybrid Quantum-Classical Computing*, Oct 27, 2022, https://doi.org/10.48550/arXiv.2210.15314. The authors distinguish between 2 classes of hybrid quantum-classical computing: (1) application agnostic focusing on the quantum computing hardware, and (2) application specific, focusing on quantum computing algorithms. Such novel technical classifications are relevant for legal classifications.

The SEA-framework can also inspire ways in which these self-regulatory and "hard laws" could be complemented by *operational instruments* derived from existing quantum use cases. The framework and its principles point towards a risk-based governance environment, with standardization, certification, production and market authorization, benchmarking, quantum quality management systems (QMS), and life cycle auditing expected to play an important role in fostering sustainable innovation. In this way, equitable access can be ensured, while putting targeted controls and guardrails in place that safely enable scalability of quantum technology, benefitting society at large.[89]

**VII. CONCLUSION AND FURTHER RESEARCH**

This article proposes a framework for Responsible QT that integrates ethical, legal, social, and policy implications into quantum R&D. The aim is to *safeguard* against risks, *engage* stakeholders, and continue *advancing* QT. The proposed SEA approach emphasizes anticipation, inclusion, reflection, and responsiveness as key dimensions of responsible research and innovation. It is aimed at researchers, developers, innovators, investors, regulators and other stakeholders who are involved in the development and commercialization of QT. Our proposed framework should be considered as a starting point for highly interdisciplinary efforts to ensure that ethical considerations are identified and discussed while QT are still shapeable. Overall, we highlight the importance of responsible innovation in the development of quantum technology to ensure its societal impact is positive.

The overarching objective of our interdisciplinary Responsible QT effort is to steer the development and use of QT in a direction not only consistent with a values-based society, but also contributive to addressing some of humankind's most pressing needs and goals. The potential global impacts of QT urge to promote responsible innovation producing responsible technologies, informed by ELSPI considerations.

Following an anticipatory approach that synchronizes precautionary measures with permissionless innovation, a first key step is the conceptualization of what *responsible* QT should mean and imply. This contribution aims to provide such an initial conceptualization by proposing the contours of a framework to balance *safeguarding, engaging,* and *advancing* QT.

We suggested to translate this framework into guiding principles for Responsible QT. Some of these principles will generally go for new and emerging technologies, and some will specifically apply to QT.[90] It is the complete set, however, that should provide guidance to quantum innovation and lead towards Responsible QT.[91] The envisioned principles should thus be understood and applied as a QT-tailored methodological framework that both complements existing frameworks for responsible innovation and translates them to the context of QT.[92] Additionally, these principles should be applicable to technological synergies such as quantum-classical interactions and quantum-AI hybrids.[93] What's more, we envision the principles to inform debates about regulatory interventions in this context, and reduce the risk of unintended counterproductive effects of such policies.

---

[88] If a general regulatory approach across QT domains is not possible or presents material challenges, a targeted approach should be pursued in the form of quantum domain, or industry specific rules. In that scenario different vertical rules would apply for quantum sensor data than for quantum computer input and output data. This would make sense from a QMS perspective with different applications having their own designated safety and security regimes, comparable to a PC having different CE marking requirements than a drone. In addition, core overarching horizontal rules could apply, analogous to universal human rights of privacy, integrity of the person, and freedom of speech, and inspired by the SEA-framework for responsible quantum innovation's principles.

[89] *See* Kop & Brongersma, *supra* note 33.

[90] For a catalogue of principles relating to quantum computing, *see* World Economic Forum, *Quantum Computing Governance Principles*, (WEF Jan. 2022) https://www.weforum.org/reports/quantum-computing-governance-principles. For broader principles pertaining to the entire suite of QT, *see* Kop, *supra* note 49.

The need for Responsible QT becomes evident when examining prospective trajectories where QT software and hardware structures are developed without adequate consideration of the various aspects of Quantum-ELSPI. We used the impact of quantum computing in information security as a case study to illustrate (1) why we need to consider Quantum-ELSPI and commit to Responsible QT, and (2) how the same ways we handled quantum computing using the proposed framework would be useful for Responsible QT more generally.[94]

Looking forward, it will be necessary to complement our proposed framework with principles for responsible quantum innovation. This will require the collaboration of multidisciplinary teams of diverse quantum stakeholders. Although QT's consequences and impact remain largely unknown, we hope this contribution may serve as an open invitation for researchers, innovators, and regulators to discuss and orchestrate normative dimensions of QT futures, and pathways to build towards them.

**ACKNOWLEDGEMENTS**
Timo Minssen's, Mateo Aboy's, and Glenn Cohen's research for this paper was supported, in part, by a Novo Nordisk Foundation Grant for a scientifically independent International Collaborative Biomedical Innovation & Law Program – Inter CeBIL (grant no. NNF17SA027784). RL thanks Mike and Ophelia Lazaridis for funding.

**One Sentence Synopsis:**
The article proposes a conceptual framework for Responsible Quantum Technology by integrating considerations about ethical, legal, social, and policy implications (ELSPI) of quantum technologies and RRI values into quantum R&D.

---

[91] Ideally, one would want to measure, benchmark, validate and certify responsible quantum technologies during their life span, denoting parameters using a data driven approach.

[92] *Compare to* E. Perrier, *The Quantum Governance Stack: Models of Governance for Quantum Information Technologies*, DIGITAL SOCIETY, QUANTUM-ELSPI TC, 1, 22, SPRINGER NATURE, (2022).

[93] *See* X. Yang, X. Chen, J. Li, X. Peng, R. Laflamme, *Hybrid quantum-classical approach to enhanced quantum metrology*. SCIENTIFIC REPORTS. 11: 672. PMID, NATURE, (2021) 33436795 DOI: 10.1038/s41598-020-80070-1

[94] We are currently moving from the NISQ Era to the Fault Tolerant Era with error corrected quantum computers consisting of 1 million stable logical qubits capable of running novel algorithms expected before 2030, featuring double exponential growth curves, potentially unlocking a new industrial revolution.

> **Box 2: Responsible Quantum Technology Summary**
>
> **WHY – Why the need for responsible quantum technology?**
> - For the last 100 years, scientists have been working on quantum science.
> - At first, quantum science focused on developing the theory of quantum mechanics to understand the principles and rules that govern physical reality at a fundamental particle level.
> - Later, these insights were applied to technology development.
> - The first quantum revolution in engineering brought us 1G QT.
> - The second quantum revolution in engineering promises 2G QT.
> - With the introduction of QT into society comes the need for ELSPI-considerations.
> - The potential impact of QT makes ELSPI-considerations ever more important. A key step is the conceptualization of what "responsible" QT means and implies.
> - This paper aims to provide such a conceptualization and suggests its operationalization by guiding principles.
>
> **WHAT – What does responsible quantum technology amount to?**
> - Proposition: the potential societal impact of QT calls for Responsible QT.
> - We should explore what "responsible" amounts to in the context of QT.
> - Analytically, Responsible QT is about integrating ELSPI-considerations into R&D processes to ensure responsible quantum innovation.
> - Normatively, Responsible QT is about minimizing harm and maximizing benefit.
> - Our proposal: Responsible QT entails an innovation process that proactively addresses risks, takes on challenges and seizes opportunities that come with the development of QT.
> - We translate this idea into the SEA-framework for Responsible QT, capturing three key aspects of Responsible QT: Safeguarding, Engaging, Advancing.
> - We propose to approach Responsible QT by the SEA-framework, responding to Responsible Research & Innovation (RRI)-dimensions.
>
> **HOW – How to pursue responsible quantum technology?**
> - Responsible QT is the aim of responsible quantum innovation.
> - The SEA-categories should be developed into a set of foundational principles to guide quantum innovation and contribute to Responsible QT.
> - This could be a mix of quantum-specific principles and generic responsible tech principles that are intrinsically relevant to QT.
> - The research community is invited to develop these guiding principles and discuss their operationalization.
> - Ideally, implementing such principles into practice should result in Responsible QT by design and default.